\documentclass[reprint,amsmath,amssymb,aps,prb,floatfix,superscriptaddress,twocolumn]{revtex4-2}

\usepackage[T1]{fontenc}
\usepackage[utf8]{inputenc}
\usepackage[english]{babel}
\usepackage{graphicx}
\usepackage{hyperref}
\usepackage{times}
\usepackage{bm}
\usepackage{placeins}
\usepackage{natbib}

\def\br{{\bf r}}
\def\bq{{\bf q}}
\def\bp{{\bf p}}
\def\bR{{\bf R}}

\def\b0{{\bf 0}}
\def\o2{{1 \over 2}}

\newcommand{\beq}{\begin{eqnarray}}
\newcommand{\eeq}{\end{eqnarray}}
\newcommand{\be}{\begin{equation}}
\newcommand{\ee}{\end{equation}}

\begin{document}
\title{Quantum Monte Carlo determination of the principal Hugoniot of deuterium.}

\author{Michele Ruggeri}
\affiliation{Maison de la Simulation, CEA, CNRS, Univ. Paris-Sud, UVSQ, Universit{\'e} Paris-Saclay, 91191 Gif-sur-Yvette, France}
\author{Markus Holzmann} 
\affiliation{Univ. Grenoble Alpes, CNRS, LPMMC, 3800 Grenoble, France}
\affiliation{Institut Laue Langevin, BP 156, F-38042 Grenoble Cedex 9, France}
\author{David M. Ceperley}
\affiliation{Department of Physics, University of Illinois, Urbana, Illinois 61801, USA}
\author{Carlo Pierleoni}
\affiliation{Maison de la Simulation, CEA, CNRS, Univ. Paris-Sud, UVSQ, Universit{\'e} Paris-Saclay, 91191 Gif-sur-Yvette, France}
\affiliation{Department of Physical and Chemical Sciences, University of L'Aquila, Via Vetoio 10, I-67010 L'Aquila, Italy}

\begin{abstract}
We present Coupled Electron-Ion Monte Carlo results for the principal Hugoniot of deuterium  
together with an accurate study of the initial reference state of shock wave experiments. 
We discuss the influence of nuclear quantum effects, thermal electronic excitations, and
the convergence of the energy potential surface by
wave function optimization within Variational Monte Carlo and Projection Quantum Monte Carlo methods.
Compared to a previous study, the
new calculations also include low pressure-temperature (P,T) conditions resulting in close agreement with experimental data, while 
our revised results at higher (P,T) conditions still predict
a more compressible Hugoniot 
than experimentally observed.
\end{abstract}

\maketitle

The determination of the hydrogen phase diagram across a large range of temperature and pressures, an important topic in planetary science,
is a challenging problem 
mainly relying on static or dynamic compression experiments or theoretical computations \cite{McMahon2012a}. 
Static compression experiments performed in diamond anvil cells are used to probe the properties of solid hydrogen at low temperature and of fluid hydrogen just above the melting conditions, while the fluid phase in a wider range of thermodynamic conditions is investigated using dynamical
compression by shock waves e.g. applying a short duration, high intensity force to a hydrogen or deuterium sample using a gas gun \cite{Holmes1995}, converging explosive shock \cite{Boriskov2005}, pulsed power \cite{Knudson2001,Knudson2017}, laser \cite{Hicks2009,Falk2012,Falk2013} or magnetically driven platforms \cite{Knudson2015}. 
Given an initial reference state, the densities and pressures of the final state of the shock (the Hugoniot)
are determined by the 
Rankine--Hugoniot relation, directly following from basic conservation laws in fluid dynamics \cite{Rankine1870}. A review of experimental techniques is presented in Ref. \cite{Nellis2006}.

The deuterium Hugoniot can be predicted from first-principle simulations based either on Density Functional Theory (DFT) or Quantum Monte Carlo (QMC) techniques.
DFT predictions appear to have good agreement with experiments \cite{Knudson2017},
which surprisingly is seemingly independent of the exchange-correlation (XC) approximation employed, while Coupled Electron Monte Carlo (CEIMC) results based on QMC techniques find a more compressible Hugoniot than obtained from experiments \cite{Tubman2015}.
An analysis of error propagation in the Hugoniot for several first-principle methods, including QMC and DFT, \cite{Clay2019} 
 showed that the insensitivity of DFT to different XC functionals
 results from important error cancellation, whereas QMC inaccuracies may get amplified in the Hugoniot predictions. In particular, the fixed node error of the electronic QMC energies
was suggested as the possible origin of the observed deviation of the CEIMC predictions
from experiments.

In this paper, we report new QMC results for the principal Hugoniot, obtained by ground state Variational Monte Carlo and Reptation Quantum Monte Carlo, within the fixed-node or fixed-phase approximation. Nuclear motion is simulated by CEIMC. Selected nuclear configurations from CEIMC trajectories are used 
to study the convergence of electronic ground state calculations. For each configuration,
the converged energy and pressure are then averaged for determining the Hugoniot.
Within this protocol, we studied systems at different densities for $T \le 8000$ K, where electronic thermal effects are negligible, and compare our computed Hugoniot to experimental data and DFT results. Particular care is given to the analysis of the cryogenic fluid reference state, which plays a fundamental role in the determination of the Hugoniot curve. 
Since we expect the fixed-node error to be less pronounced in the molecular phase at low compression, we mainly focus on the low compression-low temperature part of the Hugoniot.

This paper is organized as follows: in section \ref{sec:hugoniot} we review the Hugoniot--Rankine relation,
reviewing the current consensus for theoretical and experimental results;
in section \ref{sec:method} we describe the computational 
methods, and the protocol followed in the present study, presenting
our results in section \ref{sec:results}. 
We start with the results for the reference state,
reporting several estimates of reference energies. Then we present our results 
under low compression including several small corrections
such as electronic thermal effect and nuclear quantum effects. The last section \ref{sec:conclusions} reports our conclusions. In the appendix we report a structural analysis along the Hugoniot and some details on the Reptation Monte Carlo (RMC) calculations.


\section{The Hugoniot--Rankine relation}
\label{sec:hugoniot}
The equation of state of a material can be determined in shock experiments. A shock is applied to a system in an initial state at a pressure $P_0$ and temperature $T_0$
with energy per atom $e_0$, and  volume per atom $v_0$. Depending on the strength of the shock, the system can reach 
final states with $e$, $P$, and $v$
determined by the Hugoniot--Rankine relation
\beq
H(v,T) &= & e(v,T) - e_0 + \frac{1}{2} \left( v-v_0\right)\left[ P(v,T)+P_0\right] = 0.
\nonumber \\
\label{eq:hugoniot}
\eeq
In order to compare with the most precise shock experiments \cite{Knudson2017,Knudson2004} we set our initial conditions to be those appropriate for liquid deuterium at $T=22$ K, $P_0 = 1.24 \cdot 10 ^{-4}$ GPa and $v_0 = 135.15 a_0^3$, $a_0$ being the Bohr radius. In the following, we will express our volumes in terms of the Wigner--Seitz radius $r_s ~a_0 = \sqrt[3]{\frac{3}{4\pi}v}$, with the initial volume corresponding to $r_s^0 = 3.18353$. We stress that a precise determination of the properties of the initial reference state is needed to define the Hugoniot; see  Eq.(\ref{eq:hugoniot}).  Inaccuracy of the reference energy $e_0$ will change the Hugoniot, especially at lower temperatures. 
A detailed description of the procedure that we used to determine the reference point can be found below in section \ref{sec:reference}.


\section{Methods} \label{sec:method}
In this study we employed first principle simulations methods based on Density Functional Theory (DFT) and on Quantum Monte Carlo (QMC) electronic energy determination within the Born-Oppenheimer approximation \cite{McMahon2012a}. The nuclear configuration space is sampled using a Metropolis Monte Carlo algorithm. Nuclei are either represented by point particles in the classical (high temperature) limit or by path integrals for quantum particles, assumed here to be distinguishable \cite{Ceperley1995}. When employing the QMC electronic solution this method is called Coupled Electron-Ion Monte Carlo (CEIMC) while when employing the DFT electronic solution we will call it Born-Oppenheimer Monte Carlo (DFT-BOMC) for similarity with the commonly used Born-Oppenheimer Molecular Dynamics (BOMD) approach.
Both electronic solutions are implemented in our CEIMC code, BOPIMC. 

CEIMC can exploit electronic QMC energies from both Variational Monte Carlo (VMC) or Reptation Monte Carlo (RMC).
Both, VMC and RMC, are based on the variational principle of quantum mechanics:
for any given Hamiltonian, the exact ground state will have the lowest energy; it is thus possible
to estimate the ground state wave function by defining a trial wave function 
$\Psi_T(\bp; \br)$ with a suitable functional form, which depends on a set of $M$ variational parameters 
$\bp = \lbrace p_1, p_2, \dots , p_M\rbrace$; $\br=\lbrace \br_1, \br_2, \dots , \br_N\rbrace$
are the $N$ electronic  coordinates. Within VMC, these variational parameters are numerically optimized by minimizing a linear combination of 
the trial energy $E_T$ defined as
\beq
E_T = \left\langle E_{loc}(\br) \right\rangle_\bp =\frac{\int d\br \left\vert \Psi_T(\bp; \br) \right\vert ^2 E_{loc}(\br)}{\int d\br\left\vert\Psi_T(\bp; \br)\right\vert ^2};
\eeq
here $\langle \cdots \rangle_\bp$ stands for the mean value of a quantity over the normalized
probability distribution $\vert \Psi (\bp; \br) \vert^2$ and the local energy is
\beq
E_{loc}(\br) = \frac{\hat{\text{H}}\Psi_T(\bp;\br)}{\Psi_T(\bp;\br)},
\eeq
and its variance.

RMC \cite{Baroni1999} is a projection technique: it consists in repeatedly applying
the imaginary time evolution operator $U(\tau) = \exp(-\tau\hat{\text{H}})$ to a trial wave function, pre--optimized via VMC, which has the effect of filtering out components coming from excited states.
As with other projection methods RMC is in principle exact for $\tau \rightarrow 0$, but for fermions it suffers from the sign problem. 
To circumvent the sign problem and the associated exponentially 
increasing cost for exact Fermion calculations,
 the fixed-node or fixed-phase approximation is used.
The resulting energy is the lowest energy consistent with the nodes or phase of the assumed trial function, hence
below the VMC energy but above the ground state energy.

When using either VMC or RMC, an accurate trial wave function is important.
In our computations we used backflow Slater--Jastrow wave functions of the form
\beq
\Psi_T(\br;\bR) = J(\br;\bR) \cdot \text{Det}_{\uparrow}\left( \phi_i(\bq_j) \right)\text{Det}_{\downarrow}\left( \phi_i(\bq_j) \right).
\label{eq:trialwf}
\eeq
Here, $\bR$ represents the set of nuclear coordinates, $J(\br;\bR)$ is a general Jastrow term, symmetrical under electron exchanges, which introduces electron 
correlations, including one-- (electron-nucleus), two-- (electron-electron) and three--particle terms. The other terms in Eq.(\ref{eq:trialwf}) 
are Slater determinants (one for each spin component), which ensure that the overall electronic trial wave 
function has the correct fermionic antisymmetry. 
The single particle states $\left\lbrace\phi_i \right\rbrace$  
are determined from a DFT computation, performed using the Quantum Espresso software \cite{QE-2009,QE-2017}. 
Instead of containing bare electronic coordinates \br, 
the orbitals in the Slater determinants contain backflow transformed coordinates
$\left\lbrace\bq_i\right\rbrace$, obtained using analytical backflow transformations plus an empirical gaussian correction \cite{Feynman56,Ceperley1987a,
Holzmann2003,Pierleoni2008}.
The use of backflow coordinates introduces corrections to the nodal surface 
defined by the DFT orbitals in the Slater determinants, which lead to a significant improvement of
the accuracy of the QMC estimates, of the order of a few mHa/atom \cite{Holzmann2003}.
In CEIMC nuclear sampling is performed by a generalized Metropolis scheme\cite{Pierleoni2006}. From a given nuclear configuration a new configuration is proposed by an {\it a priori} transition probability easy to sample and an acceptance test is performed to accept or reject the move. Since the energy difference between those configurations is obtained by QMC, we need in principle to perform a new trial function optimization and a subsequent VMC or RMC energy calculation at each new attempted configuration before accepting/rejecting the move. However, this is not feasible to sample configuration space. 

In our protocol, as described in detail in the Supplemental Material of Ref. \cite{Pierleoni2016}, we usually apply several shortcuts, specific to hydrogen, to reduce drastically the computational demands. First, we rely on an accurate trial wave function for hydrogen comprising an analytic function free of variational parameters \cite{Holzmann2003}. This form is already rather accurate, in particular for a metallic system \cite{Pierleoni2008}. To this form we add empirical terms for both the Jastrow (both 1--, 2-- and 3--body) and the backflow which introduce up to 13 variational parameters. It is important to include these terms since they lower the energy by about 1 mHa/atom and reduce the variance by $40\%$ for states of hydrogen around the dissociation transition.  Rather than performing the parameter optimization for each proposed nuclear configuration at each density, we select a number of statistically independent nuclear configurations (few tens), generated with an unoptimized trial function. We then optimize the variational parameters for each configuration. We then use the average values for subsequent generation of the nuclear trajectory. This procedure takes into account the variation of the parameters with density but the variational parameters are not tailored for the specific configuration. The bias in the energy is less than the statistical precision of the the CEIMC method \cite{Pierleoni2016}. 
Because CEIMC is based on sampling the Boltzmann distribution, any bias introduced by inaccurate solutions of the electronic problem can be corrected for by reweighting the nuclear configurations \cite{Liberatore2011}. A third shortcut is that we do not perform RMC calculations of the electronic energy to advance the nuclear sampling but we run with VMC energy and we estimate the improvement from VMC to RMC on a selected subset of nuclear configurations. 
As reported in the appendix, corrections are negligibly small, within the statistical accuracy of our results, justifying our approach {\em a posteriori}.

One further limitation of our methods is the assumption of ground state electrons. From previous works on the deuterium Hugoniot by BOMD \cite{Knudson2017} it is known that electronic thermal effects can be relevant at higher temperatures/compressions. In order to assess the relevance of those effects we performed a DFT-BOMC study of the deuterium Hugoniot employing the vdW-DF1 XC approximation within ground state DFT to be compared with the results of Ref. \cite{Knudson2017} where, among others, the same XC approximation, but with thermal electrons based
on Mermin functional approach, has been employed. As discussed below,
we infer that along the Hugoniot line, electronic thermal effects make a small corrections to the EOS: the pressure difference at $\sim 10000K$ is $\simeq 6\%$ and becomes negligible at lower compressions and temperatures. We therefore expect similar small effects
within a QMC framework and negligible bias due to thermal electronic effects in CEIMC below 10000K.
By performing excited state VMC calculations for selected fixed nuclear configurations at the highest temperature as described in section \ref{sec:thermal_electrons} we explicitly confirm the minor
role of electronic temperature effects.

In most of our calculations we model nuclei as classical point particles. A noticeable exception is at the reference point (initial state) where nuclear quantum effects are so large to ensure a liquid rather than a crystalline state. To establish the physical properties of the reference state we employ three different models of increasing accuracy to be discussed in section \ref{sec:reference}. Nuclear quantum effects for states along the Hugoniot are small because of the high temperature. At the lowest temperature considered (2000K) nuclear quantum corrections on the Hugoniot are significant since the variation of the Hugoniot function with density is small. In this case we estimate NQE by running DFT-BOMC with quantum nuclei represented by Path Integrals (see section \ref{sec:NQE}).    

\section{Results}
\label{sec:results}
\subsection{The reference point}
\label{sec:reference}
When determining the Hugoniot curve described in Eq.(\ref{eq:hugoniot}) it is important to have a reliable
estimate of the properties of the reference state.
In this work we assumed  a density of
$\rho_0 = 0.167$ g/cm$^3$, corresponding to an atomic volume of $v_0 = 135.15 \,a_0^3$ or $r_s^0 = 3.18353$, as in the experiments \cite{Knudson2017,Knudson2004}. This corresponds to liquid deuterium at a pressure of 124KPa at a temperature of 22 K. 

There is an ambiguity in determining the energy of the reference point in an approximate computational method.  One might expect that numerical errors in the energy difference $e(v,T)-e(v_0,T_0)$ would cancel in determining the $H$ function of Eq. (1). Note that the individual energies are about $0.5$ Ha/atom but we need an accuracy in the difference of roughly $10^{-4}$ Ha/atom  However at high pressure, (up to 5 fold compression) there is a large change in the state of the molecules, and it is unlikely there would be a full cancellation of errors in an approximate method.  

At low pressures, CEIMC is a very precise method because the two most important sources of errors, the fixed-node approximation and finite size effects, are very small at low density.  This is because in a system with a large gap, electrons are well localized and the electrons within the same molecule have different spins so do not need to be antisymmetrized. Also the exchange effects between molecules are small and those effects are well approximated by the Slater-Jastrow trial function. 

Rather than using the CEIMC method to directly estimate $(e_0,P_0)$ we have used several indirect approaches which are based on the idea that at low density the system consists of weakly interacting deuterium molecules. CEIMC would require a large number of imaginary time slice in the path integral representation of the deuterons, resulting in an exceedingly slow dynamics to explore the phase space of the disordered molecules.
In earlier work we used energies from the properties of an isolated deuterium molecule with experimental information on how D$_2$ molecules interact at low temperatures and densities to determine the reference point energy, pressure and volume.  
We arrived at an estimate of $r_{s0}=3.18353$, $e_0=-0.583725$ Ha/atom  and $P_0=4.2 \times 10^{-9} a.u.=1.2 \times 10^{-4} GPa$

For a new estimates we note that at the reference point the typical distance between molecules is about 4 times the molecular bond length. Thus, we first consider a system of quantum molecules interacting through the Silvera--Goldmann potential\cite{Silvera1978}. This is a spherically symmetric potential between molecular centres of mass. 
For this model we performed PIMC simulations using the pair action \cite{Ceperley1995} of a system with $N=32$ molecules. 
Convergence of the results with the imaginary time step was investigated by performing simulation with an increasing number of time slices. It was found that 4 time slices were enough for the convergence of the energy at 22K. 
As expected we found a liquid hydrogen molecular state. 
To estimate the internal energy and presssure  
we included a tail correction for the interaction outside of the periodic box.

To further test the CEIMC procedure, we generated the bond distances and angles of the molecules using a second PIMC simulation with the Kolos--Wolniewicz  (KW) potential\cite{Kolos1964}. 
A new estimate of the reference point properties, obtained by adding SG and KW energies,  is reported in Tab.\ref{tab:effective}.

\begin{table}[tpb]
\caption{Total, kinetic energy per atom and pressure computed using the Silvera--Goldman (SG) 
for the molecular liquid and Kolos--Wolniewicz (KW) effective potentials for the isolated molecule. 
}
\label{tab:effective}\begin{tabular}{cccc}
\hline
                       &         SG    &    KW    & Total \\
\hline                                            
$e_{tot}$ (Ha/at.)  & $-2.225(2) \times 10^{-4}$ & -0.583598(2) & -0.583821(2) \\
$e_{kin}$ (Ha/at.)  & $ 8.893(3) \times 10^{-5}$  & 0.001701(3) &  0.001790(3) \\
$P$ (GPa)    &   -0.0050(1)          & 0  &       -0.0050(1)   \\

\hline
\end{tabular}
\end{table}

To obtain another estimate of the reference point properties we build nuclear configurations as follows. From the trajectory of the SG model we extracted 120 molecular configurations, representing the position of the center of mass of the D$_2$ molecules. From the PI simulations of the KW potential, we extract $32 \times 120$ snapshots that we assigned randomly to the 32 molecules in the different configurations to obtain 120 atomic configurations with 64 nuclei each.
These configurations are used as source to calculate the Coulomb potentials in both the electronic QMC and DFT calculations; in QMC for each configuration we optimized the trial wave function using the correlated sampling scheme, then we computed energy and pressure with both VMC and RMC. Finally we averaged over the configurations and added size corrections\cite{Holzmann2016}, reported in Tab. \ref{tab:size_effects}, to obtain the properties of the reference state. Results are reported in table \ref{tab:e0p0} and compared to the SG+KW model. Table \ref{tab:e0p0} also reports the reference energy used previously \cite{Tubman2015,Clay2019}, based on the energy of a single D$_2$ molecule \cite{Kolos1964}, the binding energy of solid D$_2$ \cite{Silvera1980} and its heat capacity \cite{Sauers1986} and the reference point values obtained with the same 120 molecular configurations using the vdW-DF1 XC functional. These values will be used in the DFT-BOMC Hugoniot.
\begin{table}[tpb]
\caption{VMC and extrapolated RMC results for the reference system with $N=64$ atoms, $T=22$ K and $r_s = 3.18353$, averaged over 120 atomic configurations. }
\begin{tabular}{ccccc}
\hline
                 &      SG+KW             &          VMC &         RMC  
                 & VdW-DF1\\
\hline                                            
E$_{el}$(Ha/at.) & -0.585611(4)         &  -0.58013(6) & -0.58570(6)  &      -0.60265(6) \\ 
E$_{tot}$(Ha/at.) & -0.583821(2)         &  -0.57828(6) & -0.58385(6)  &   -0.60081(6) \\ 
P (GPa)          & -0.0050(1)             &      0.52(6) & -0.17(6)     &     -0.54(4) \\ 
\hline
\end{tabular}
\label{tab:e0p0}
\end{table}

The reference energies for different models are different. 
The RMC energy is lower than VMC energy by $5.57$ mHa/at. But the SG+KW energy is only $0.1$ mHa/at. higher, almost within error bars than the RMC result. Also the value used in previous work \cite{Tubman2015} is only $0.13$mHa/at. higher than the reference RMC energy. 
The vdW-DF1 has a lower energy than the RMC reference but this depends on energy cutoff and pseudopotential: we employed a scalar relativistic PAW pseudopotential with an energy cutoff of $40$ Ry. In computing the Hugoniot within the VMC or DFT models it is important to use the reference point of that model since we expect some cancellation of errors. In both QMC and DFT we did not consider the pressure of the reference point which in all cases is comparable to the statistical accuracy of the pressure entering the Hugoniot function (see below).  For the RMC calculations we find the uncertainty in the reference point energy is about $0.1$ mH/atom, and so does not affect our final estimate of the Hugoniot: see below. 

\begin{center}
\begin{figure}
\includegraphics[angle=0,width=\columnwidth]{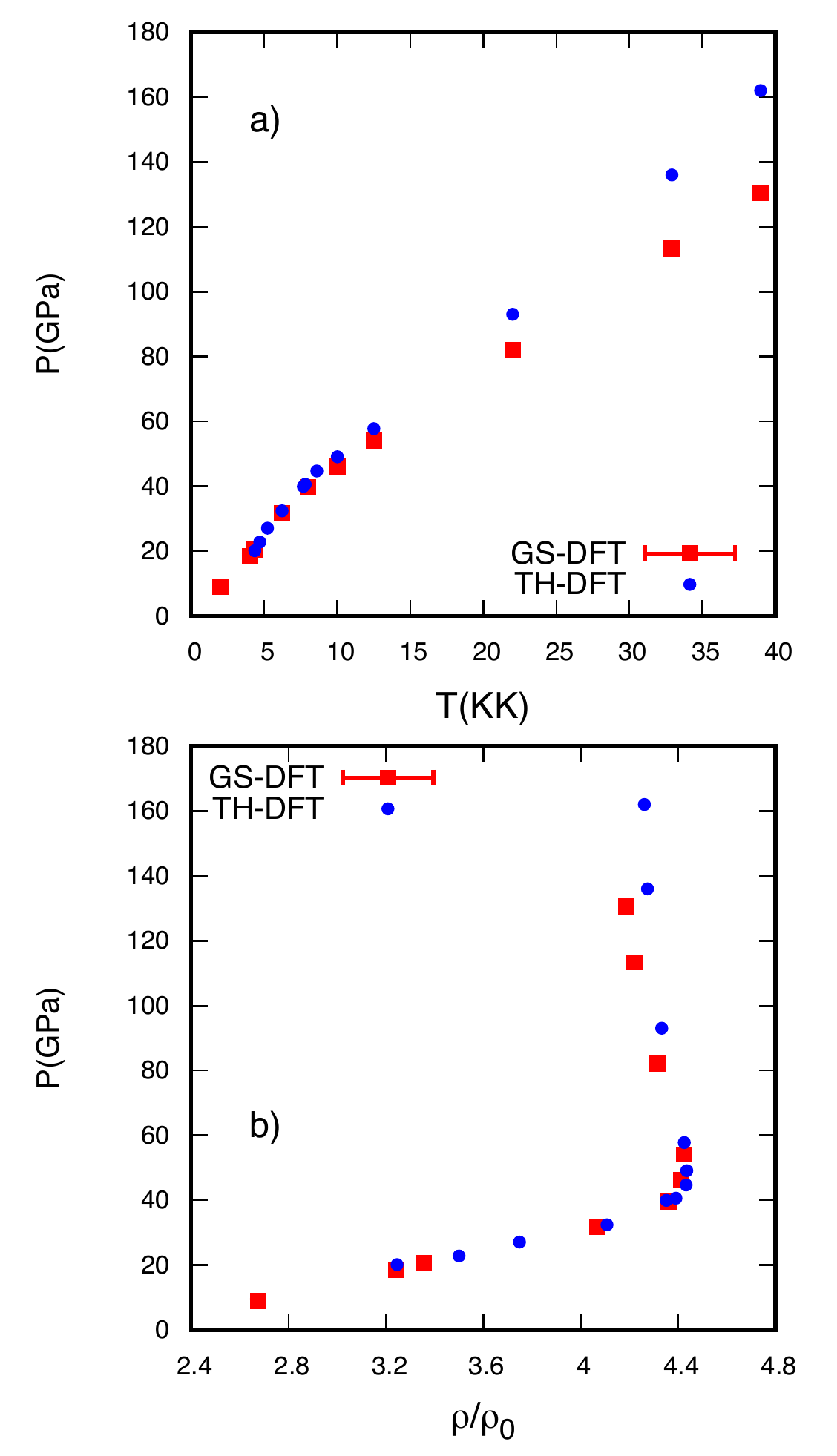}
\caption{Comparison between ground state (red squares, this work) and thermal electrons (blue circles, \cite{Knudson2017}) for thermodynamic points along the principal Hugoniot as obtained with DFT-vdW-DF1 based simulations. None of the points included nuclear quantum effects in the 
calculations.}
\label{fig:dft_df1}
\end{figure}
\end{center}

\begin{center}
\begin{table*}
\caption{VMC and RMC results for energy per atom, pressure and Hugoniot. The values include finite size effects and nuclear quantum and thermal effects.}
\label{tab:results}\begin{tabular}{ccc|cccc|cccc}
\hline
$T$ (K) & $r_s$ & $\rho / \rho_0$& E$_{VMC}$(Ha/at.) & P$_{VMC}$ (a.u.) & P$_{VMC}$(GPa) & H$_{VMC}$(Ha/at.) & E$_{RMC}$(Ha/at.) & P$_{RMC}$(a.u.) & P$_{RMC}$(GPa) & H$_{RMC}$(Ha/at.)\\
\hline
22   & 3.18353 & 1.00 & -0.57828(6) & 0.000018(2) & 0.52(6)  &            & -0.58385(6) & -0.000006(2) & -0.17(6)& \\
\hline

2000 & 2.20 & 3.03 & -0.56845(7) & 0.000337(5) & 9.9(1) & -0.0055(3) & -0.57192(5) & 0.000308(4) & 9.1(1) & -0.0021(2) \\
     & 2.25 & 2.83 & -0.56921(6) & 0.000270(4) & 7.9(1) & -0.0028(2) & -0.57270(5) & 0.000244(4) & 7.2(1) & 0.0004(2) \\
     & 2.30 & 2.65 & -0.56933(7) & 0.000243(4) & 7.1(1) & -0.0014(2) & -0.57296(6) & 0.000229(4) & 6.7(1) & 0.0012(2) \\
     & 2.40 & 2.33 & -0.56962(7) & 0.000209(5) & 6.1(1) &  0.0005(2) & -0.57371(5) & 0.000176(4) & 5.2(1) & 0.0033(2) \\

\hline
4000 & 1.80 & 5.53 & -0.5424(2) & 0.001611(14) & 47.4(4) & -0.0534(8) & -0.5451 (2) & 0.001599(13) & 47.1(4) & -0.0499(7) \\
     & 2.00 & 4.03 & -0.5503(1) & 0.000826(9)  & 24.3(3) & -0.0141(5) & -0.5536 (1) & 0.000808(8) & 23.8(2) & -0.0109(4) \\
     & 2.20 & 3.03 & -0.5541(1) & 0.000461(6)  & 13.6(2) &  0.0032(3) & -0.5581 (1) & 0.000443(6) & 13.0(2) & 0.0056(3) \\

\hline
8000 & 1.80 & 5.53 & -0.5075(2) & 0.001832(16) & 53.9(5) & -0.0308(9) & -0.5106(2) & 0.001770(15) & 52.1(4) & -0.0249(9) \\
     & 1.85 & 5.10 & -0.5085(2) & 0.001533(16) & 45.1(5) & -0.0136(9) & -0.5120(2) & 0.001475(16) & 43.4(4) & -0.0083(9) \\
     & 1.90 & 4.70 & -0.5098(2) & 0.001295(14) & 38.1(4) & -0.0006(8) & -0.5136(2) & 0.001236(13) & 36.4(5) & 0.0044(7) \\
     & 1.92 & 4.56 & -0.5100(2) & 0.001259(14) & 37.1(4) &  0.0017(8) & -0.5140(2) & 0.001203(14) & 35.4(4) & 0.0063(8) \\
     & 2.00 & 4.03 & -0.5141(3) & 0.001045(17) & 30.7(5) &  0.0110(9) & -0.5183(3) & 0.000991(16) & 29.2(5) & 0.0151(9) \\

\hline
\end{tabular}
\end{table*}
\end{center}

\subsection{Hugoniot computations}
We performed both DFT-BOMC and CEIMC calculations for  54 classical deuterons in a periodic box. In both DFT and CEIMC we sum over a $4\times 4\times 4$ regular grid of k-points. Single-electron orbitals in the CEIMC trial function are from PBE-DFT \cite{Perdew1996} with an plane wave energy cutoff of 40 Ry while in DFT-BOMC we employed vdW-DF1 xc approximation wiuth the same energy cutoff. 

Using DFT-BOMC we ran at temperatures of 2000K, 4000K, 4446K, 6207K, 8000K, 10000K, 12500K, 22000K, 32900K, and 39000K to compare directly with results of Ref. \cite{Knudson2017} but using ground state electrons. In Fig. \ref{fig:dft_df1} we report this comparison for the EOS along the principal Hugoniot. We observe a good agreement below  $T=10000K$ while above this temperature electronic thermal effects become increasingly large. Note that our small simulation cells do not show (nuclear) finite size effects invoked in Ref. \cite{Knudson2017} as possible source of inaccuracy in the CEIMC results of Ref. \cite{Tubman2015}.

With CEIMC we performed new computations beyond those reported in Ref. \cite{Tubman2015}. In particular we ran a lower temperature isotherm at $T=2000K$ and more computations at $T=4000K$ and $T=8000K$. We did not consider higher temperatures to avoid the need of considering thermal electrons (see subsection \ref{sec:thermal_electrons} for a discussion).  
At each thermodynamic point, after the generation of the trajectory we selected about 100 nuclear configurations for which we optimized the trial function individually to get the VMC estimates and to perform RMC analysis. In addition we considered both single-- and two-body finite size effects as described in Ref. \cite{Holzmann2016}. 
The pressure was estimated using the virial estimator
\beq
P = \frac{2 e-u}{4 \pi r^3_s}
\label{eq:pressure}
\eeq
where $e$ is the total energy per atom and $u$ is the potential energy per atom. For each temperature and density we compute the Rankine--Hugoniot function Eq.(\ref{eq:hugoniot}), and we find the Hugoniot point, i.e. $(v,T): H(v,T)=0$, using a linear interpolation.
Results of our analysis, including finite size effects, are reported in table \ref{tab:results} while finite size effects on energy and pressure are reported in table \ref{tab:size_effects}. Note that ``electronic'' size
corrections are important to the final result, and increase with the temperature of the system, as the system becomes more metallic. 

\begin{figure}
\includegraphics[width=\columnwidth]{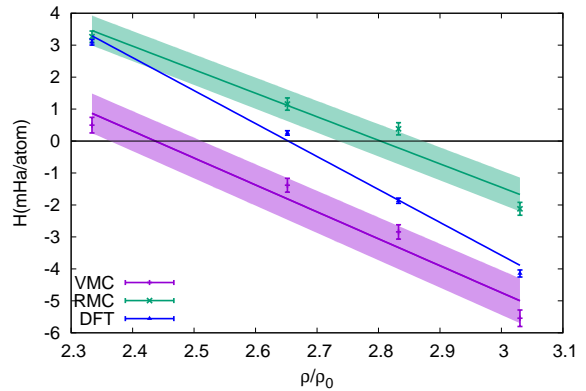}
\includegraphics[width=\columnwidth]{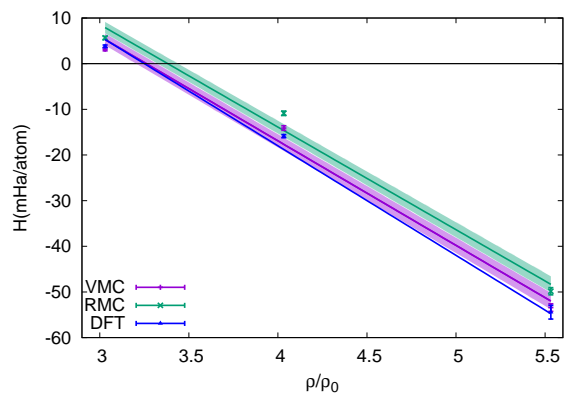}
\includegraphics[width=\columnwidth]{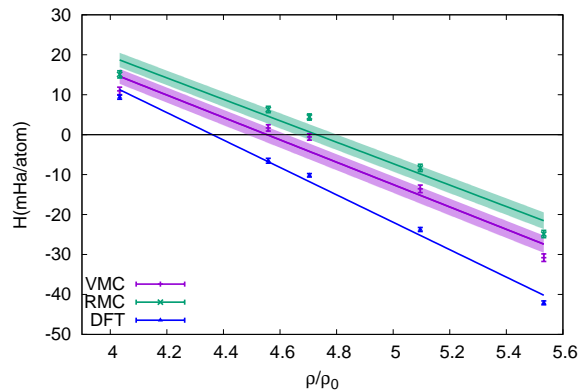}
\caption{QMC Hugoniot for $T=2000$, 4000 and 8000 as a function of the ratio between density $\rho$ and density at the reference point $\rho_0$. The shaded regions represent the uncertainties in the interpolations of the QMC data. Uncertainties due to changes in the reference energy are within these regions.}
\label{fig:hugovsT}
\end{figure}

Figure \ref{fig:hugovsT} reports $H(T,v)$ for the three isotherms. Each panel show results for VMC and RMC and for the DFT-BOMC calculations.


Before summarising our results in section \ref{sec:summary}, in the next two subsections we need to address two effects.

\subsection{Electronic thermal effects}
\label{sec:thermal_electrons}
A way to establish the relevance of electronic thermal effects in QMC, and possibly to estimate their size, is to perform, for given  nuclear configurations, electronic QMC calculations with excitations built into the trial function. At the VMC level this is not difficult and can be realized by considering excited Slater determinants in the trial function. With enough excitations, thermal averages can be obtained by weighting each contribution with its Boltzmann weight $\sim e^{-\beta (E_n-E_0)}$. Here $E_0$ is the ground state energy and $E_n$ the energy of the $n$--th excited state. Unfortunately the number of relevant excitations increases rapidly with temperature, in particular near metallization where the energy gap is small. It also depends on the grid of twists, so computing in practice the thermal correction is impractical. What we can do is to compare how the QMC energy and pressure for each excitation differ from the corresponding DFT ones. If the differences are small we can expect that electronic thermal effects in QMC are similar and of the same size as the ones in the DFT calculations. 

\begin{figure}
\includegraphics[width=\columnwidth]{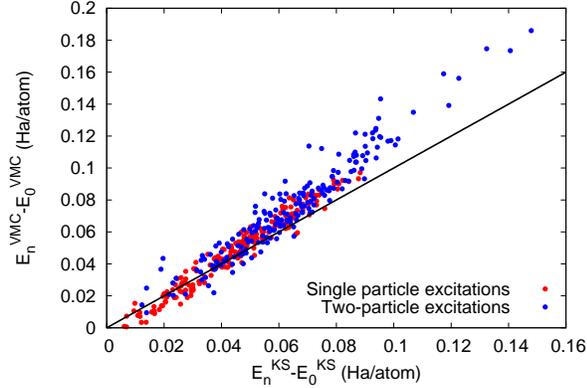}

\includegraphics[width=\columnwidth]{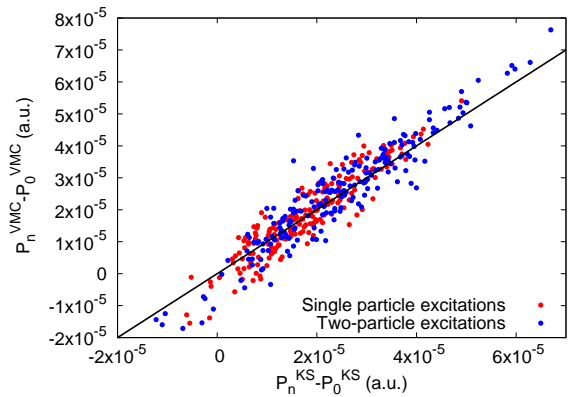}

\caption{VMC excitation energies versus Kohn--Sham excitation (top) and difference in pressure from VMC and DFT computations of excited states (bottom). The black line indicates equality between the VMC and DFT estimates.
}
\label{fig:T_effects}
\end{figure}

We selected several nuclear configurations generated during the CEIMC sampling at $T=8000$ K and $r_s = 1.88$, very close to the density of the principal Hugoniot at this temperature, and for each configurations we ran VMC with single and two particle excitations. The trial wave functions for the excited states were obtained including single and double particle excitations in the Slater determinants in Eq.(\ref{eq:trialwf}). Backflow transformation were used, but the trial wavefunctions were not re--optimized independently for each excitation.
In figure \ref{fig:T_effects} we compare energy and pressure from VMC excited calculations and DFT excited calculations (in fact the eigenvalues of the Kohn-Sham solution). 
Even if the values of the energy and pressure estimated with DFT and QMC can differ, the energy (pressure) differences between excited and ground states display a very similar behaviour. Since the effect of finite temperature is to increase the population of excited states with their relative Boltzmann weight (w.r.t. the ground state), having the same excitation energies means having the same thermodynamic properties. From the bottom panel of Fig. \ref{fig:T_effects} we see that the virial estimator for the pressure displays the same behaviour; this behaviour holding for energy and pressure means that the same will apply to the Hugoniot, Eq.(\ref{eq:hugoniot}).
Since we have shown that at $T\leq 8000$ thermal effects are negligible in DFT, we conclude that we can safely use ground state QMC methods to describe the electrons in our system for $T\leq 8000K$.

\begin{table}[htpb]
\caption{Finite size effects on the energy and pressure of a system of deuterium at different temperatures and densities; the system is made of $N=64$ atoms for the $T=22$ K reference state and $N=54$ atoms in the other cases.}
\label{tab:size_effects}\begin{center}
\begin{tabular}{cccc}
\hline
$T$ (K) & $r_s$ & $\Delta E$ (Ha/at.) & $\Delta P$ (a.u.) \\
\hline
22   & 3.18353 & 0.00050 & 0.0000012  \\
\hline
2000 & 2.20    & 0.00156 & 0.0000104  \\
     & 2.25    & 0.00145 & 0.0000088  \\
     & 2.30    & 0.00137 & 0.0000078  \\
     & 2.40    & 0.00119 & 0.0000057  \\
\hline
4000 & 1.80    & 0.00326 & 0.0000492  \\
     & 2.00    & 0.00259 & 0.0000277  \\
     & 2.20    & 0.00172 & 0.0000123  \\
\hline
8000 & 1.80    & 0.00423 & 0.0000703  \\
     & 1.85    & 0.00395 & 0.0000597  \\
     & 1.90    & 0.00383 & 0.0000543  \\
     & 1.92    & 0.00373 & 0.0000504  \\
     & 2.00    & 0.00316 & 0.0000371  \\
\hline
\end{tabular}
\end{center}
\end{table}

\subsection{Nuclear Quantum effects}
\label{sec:NQE}
In the results described up to now, we have neglected nuclear quantum effects (NQE) except at the reference point.  NQE in high pressure hydrogen and deuterium could be relevant, depending on the thermodynamic conditions, because of the light nuclear mass and the molecular character. The energy of an isolated D$_2$ molecule is higher by 1089K/atom over a molecule of classical ions, a contribution that is relevant for the present level of accuracy. We also note that the zero temperature, zero pressure atomic volume of solid H$_2$ is 16\% higher than that of solid D$_2$, showing the significance of nuclear quantum effects in the equation of state.
To gauge the influence of nuclear quantum effects on our results we performed Path Integral simulations of deuterium at $T=2000K$ at $r_s = 2.20$, $2.25$ and $2.30$. Because of the very good structural agreement observed between CEIMC and DFT-BOMC for classical nuclei (see appendix \ref{sec:appendix1}) we ran these calculation using DFT-BOMC since that requires a smaller computational resource. 
Path Integral simulations were performed with the strategy implemented in CEIMC and detailed in Refs \cite{Pierleoni2006,Pierleoni2016} but using DFT to evaluate the electronic energy. We found that 4 nuclear time slices were enough to converge the time step. Comparing results of these calculations to the results for classical nuclei we compute the correction to the energy, pressure and Hugoniot. The same corrections are used for both the CEIMC and DFT results.

\begin{figure}
    \centering
    \includegraphics[width=\columnwidth]{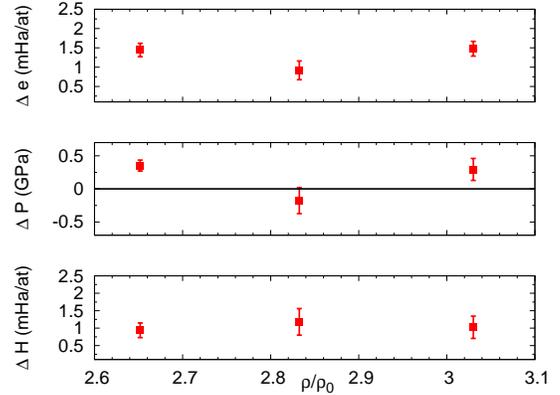}
    \caption{Difference in energy per atom, pressure and Hugoniot between a classical and a quantum system of 54 deuterium atoms at $T=2000$ K at different densities.}
    \label{fig:quantum_effects}
\end{figure}

In Fig.\ref{fig:quantum_effects} we report the changes in total energy, pressure and $H(T,v)$. Corrections are small (in particular on the pressure) but on the scale of the Hugoniot (see panel a) of fig. \ref{fig:hugovsT}) a correction of $\sim 1$ mHa/atom is significant. At higher temperatures the NQE is unimportant to the Hugoniot. 

\subsection{Summary of Hugoniot results}
\label{sec:summary}
\begin{figure*}
\includegraphics[width=\textwidth]{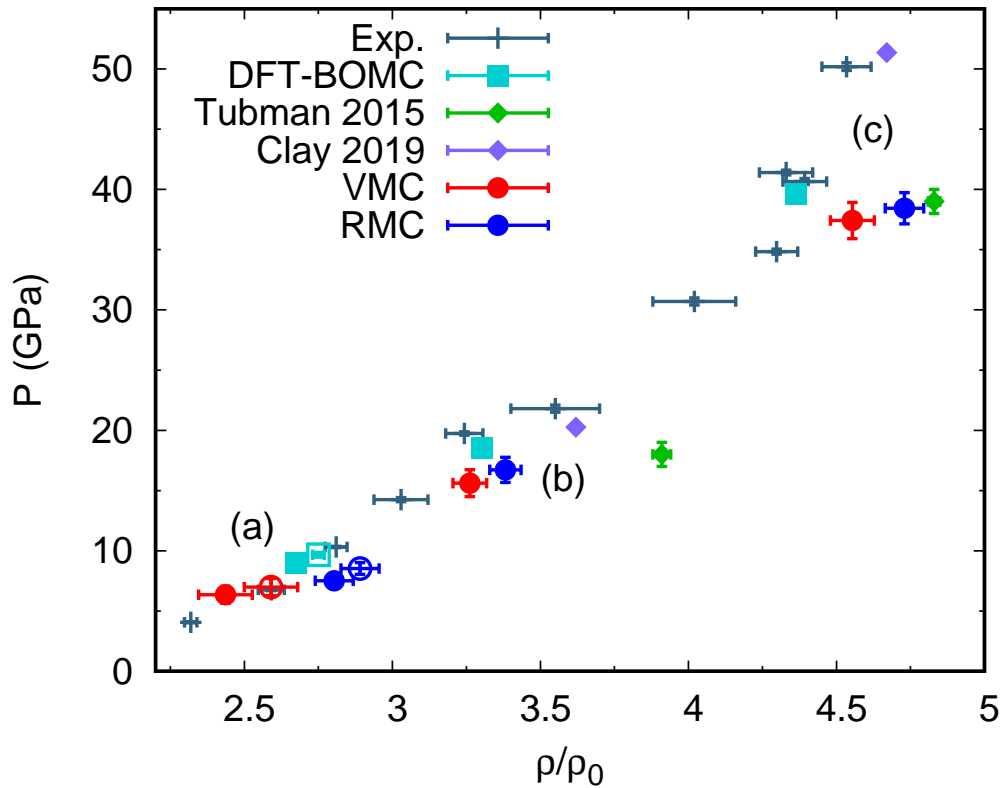}
\caption{QMC Hugoniot compared with previous results. We show experimental data \cite{Knudson2015,Knudson2017} 
and theoretical DFT-BOMC and QMC data \cite{Tubman2015, Clay2019}; (a), (b) and (c) label different data sets, corresponding to computations performed at different temperatures, with $T=2000$, $4000$ and $8000$ K respectively. Open symbols at $T = 2000$ K represent our QMC and DFT-BOMC results including nuclear quantum effects (NQE), all other results are obtained with classical nuclei.}
\label{fig:hugo}
\end{figure*}

We show in figure Fig.\ref{fig:hugo} our computed Hugoniot curve, using the new values for the reference point, and we compare with experimental data and previous simulations. The figure also reports our DFT-BOMC points already shown in figure \ref{fig:dft_df1}. VMC and RMC data are rather close to experimental determinations of the Hugoniot line, in particular at low compression and temperature while the deviation is larger at higher compression near the stiffening of the Hugoniot. This stiffening appears when the molecular fraction is small  and the nearly fully atomic system becomes less compressible. Our structural analysis, reported in the appendix, shows a molecular fraction of 10-15\%  in the system at 8000K along the principal Hugoniot. Conversely at low compression and temperature the system is fully molecular and its electronic configuration (molecular-singlet) should be rather simple and easy to model by QMC. Agreement of our results with experimental data is particularly good at T=2000K, even better for the VMC estimate than for the RMC one. One cannot exclude the possibility of a residual error in the correction of the RMC results for time step and projection time errors. NQE effects are small and shift the points to higher compression and pressure. 
In general, our deuterium model appears to be slightly more compressible than experiment with the VMC description being slightly less compressible than the RMC model. This is in line with the trend observed in the previous CEIMC determination of the Hugoniot \cite{Tubman2015} but our new results are closer to the experimental curve, in particular at low compression. Figure \ref{fig:hugo} also reports the previous CEIMC prediction \cite{Tubman2015} together with its corrected values
from Ref. \cite{Clay2019}. Compared to the experimental data, the quality of our present predictions of the Hugoniot at 4000K and 8000K is equivalent to Clay's et al. \cite{Clay2019} fixed note error corrections. However, we note that these points do not agree in terms of absolute compression and pressure.

\section{Conclusions}
\label{sec:conclusions}
We have reported a new investigation of the deuterium principal Hugoniot by CEIMC methods. By combining several QMC methods, VMC and RMC, we focused on the low compression-low temperature part of the Hugoniot where deuterium is  molecular. We also performed a careful study of the cryogenic reference state using both effective model potentials with Path Integral Monte Carlo and electronic QMC calculations. We obtained a good agreement with the low compression experimental Hugoniot while at the higher temperature our predicted Hugoniot remains slightly more compressible than experiment\cite{Knudson2017}. Our Hugoniot is essentially in agreement with the one from recent first-principle analysis \cite{Clay2019} of previous CEIMC calculation \cite{Tubman2015}. However comparing predictions at fixed temperature our Hugoniot point has a smaller compression and pressure than from Ref. \cite{Clay2019}, hence the two models have different Equation of States. The origin of this deviation is unclear and deserves further investigation.

\begin{acknowledgments}
This work has received funding from the European Union’s Horizon 2020 research and innovation program under the grant agreement No. 676531 (project E-CAM) and the grant agreement No 824158 (project EoCoE 2).
D.M.C. was supported by DOE Grant NA DE-NA0001789 and by the Fondation NanoSciences (Grenoble). C.P. was supported by the Agence Nationale de la Recherche (ANR) France, under the program ``Accueil de Chercheurs de Haut Niveau'' project: HyLightExtreme. Computer time was provided by the PRACE Project 2016143296, ISCRAB (IsB17\_MMCRHY) computer allocation at CINECA Italy, the high-performance computer resources from Grand Equipement National de Calcul Intensif (GENCI) Allocations 2018-A0030910282 and 2019-A0070910282, by an allocation of the Blue Waters sustained petascale computing project, supported by the National Science Foundation (Award OCI 07- 25070) and the State of Illinois, and by the Froggy platform of CIMENT, Grenoble (Rh{\^o}ne-Alpes CPER07-13 CIRA and ANR-10-EQPX-29-01).
\end{acknowledgments}


\bibliographystyle{apsrev4-2}
\bibliography{hugoniot}

\newpage

\appendix
\section{Structural analysis}
\label{sec:appendix1}
\begin{figure}
\includegraphics[angle=0,width=\columnwidth]{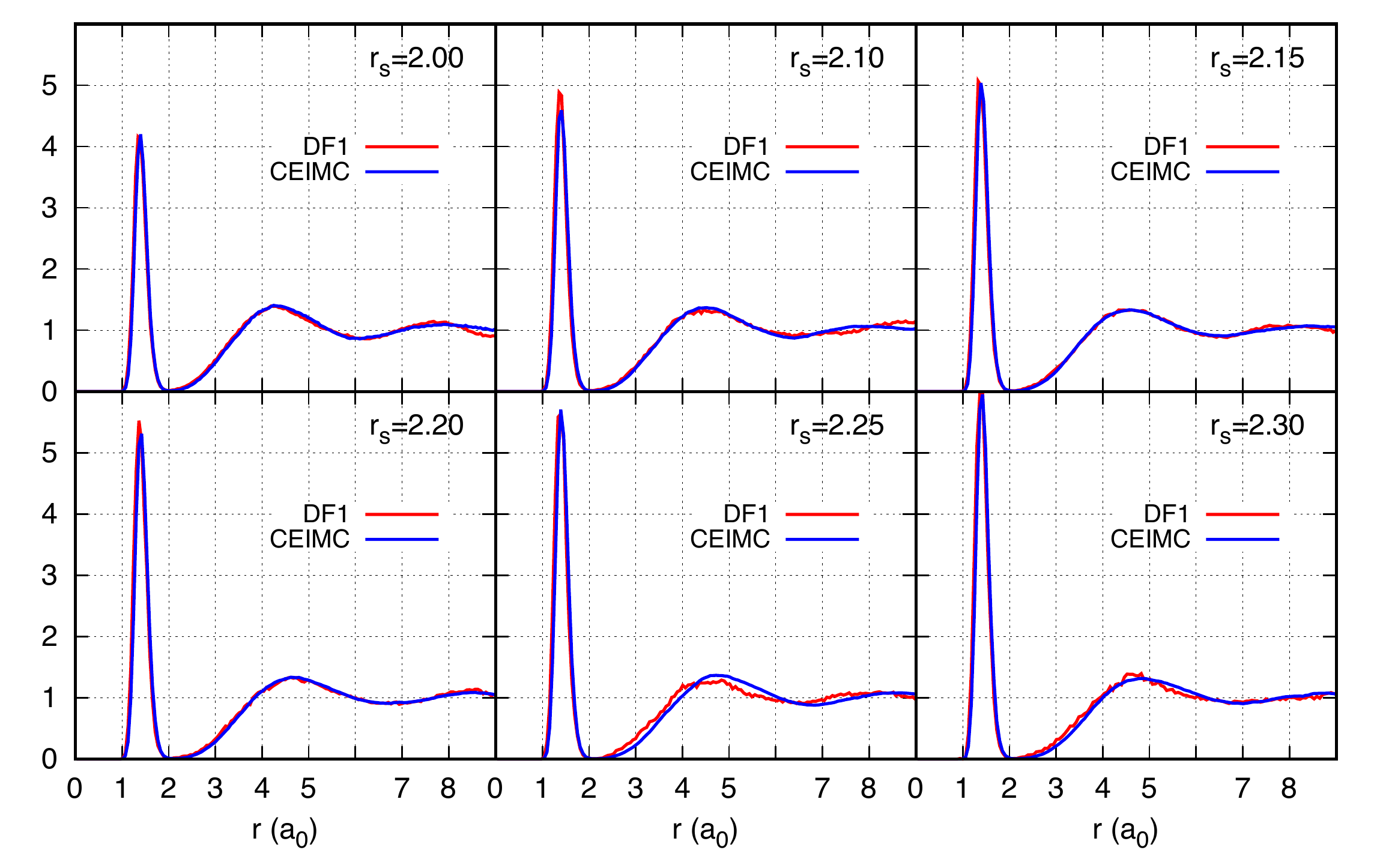}
\caption{Proton-proton pair correlation function from CEIMC and from DFT-BOMC along the T=2000K isotherm.}
\label{fig:2TKKgr}
\end{figure}

\begin{figure}
\includegraphics[angle=0,width=0.7\columnwidth]{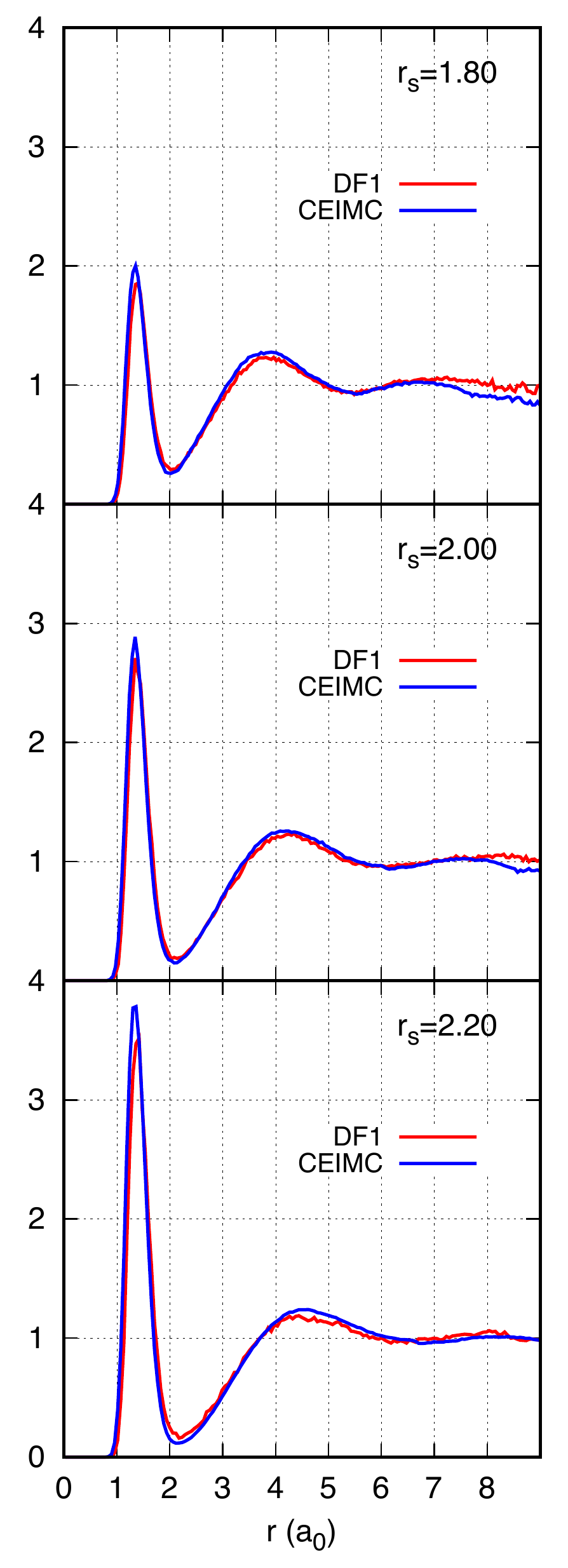}
\caption{Proton-proton pair correlation function from CEIMC and from DFT-BOMC along the T=4000K isotherm.}
\label{fig:4TKKgr}
\end{figure}

\begin{figure}
\includegraphics[angle=0,width=\columnwidth]{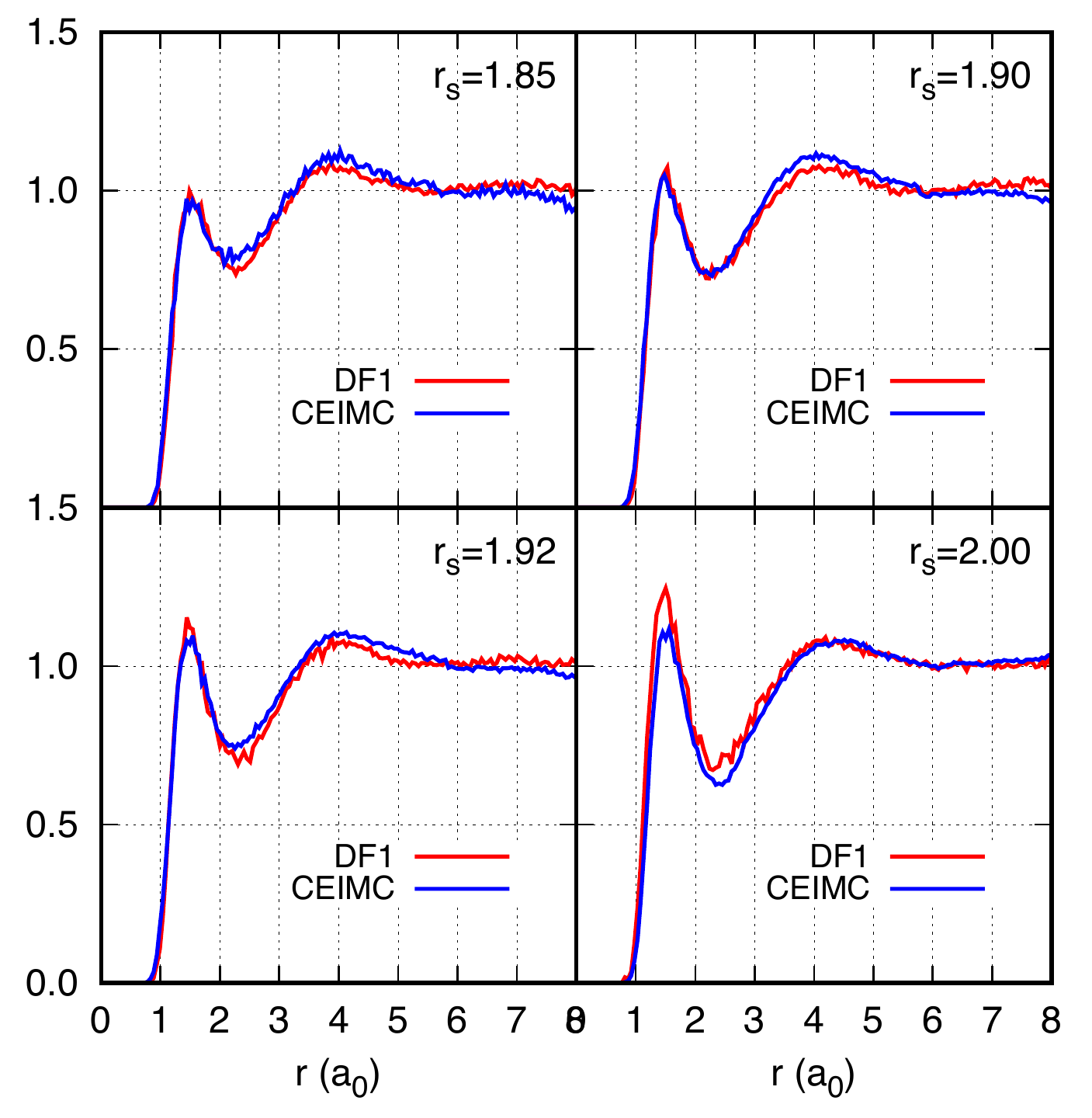}
\caption{Proton-proton pair correlation function from CEIMC and from DFT-BOMC along the T=8000K isotherm.}
\label{fig:8TKKgr}
\end{figure}

\begin{figure}
\includegraphics[angle=0,width=\columnwidth]{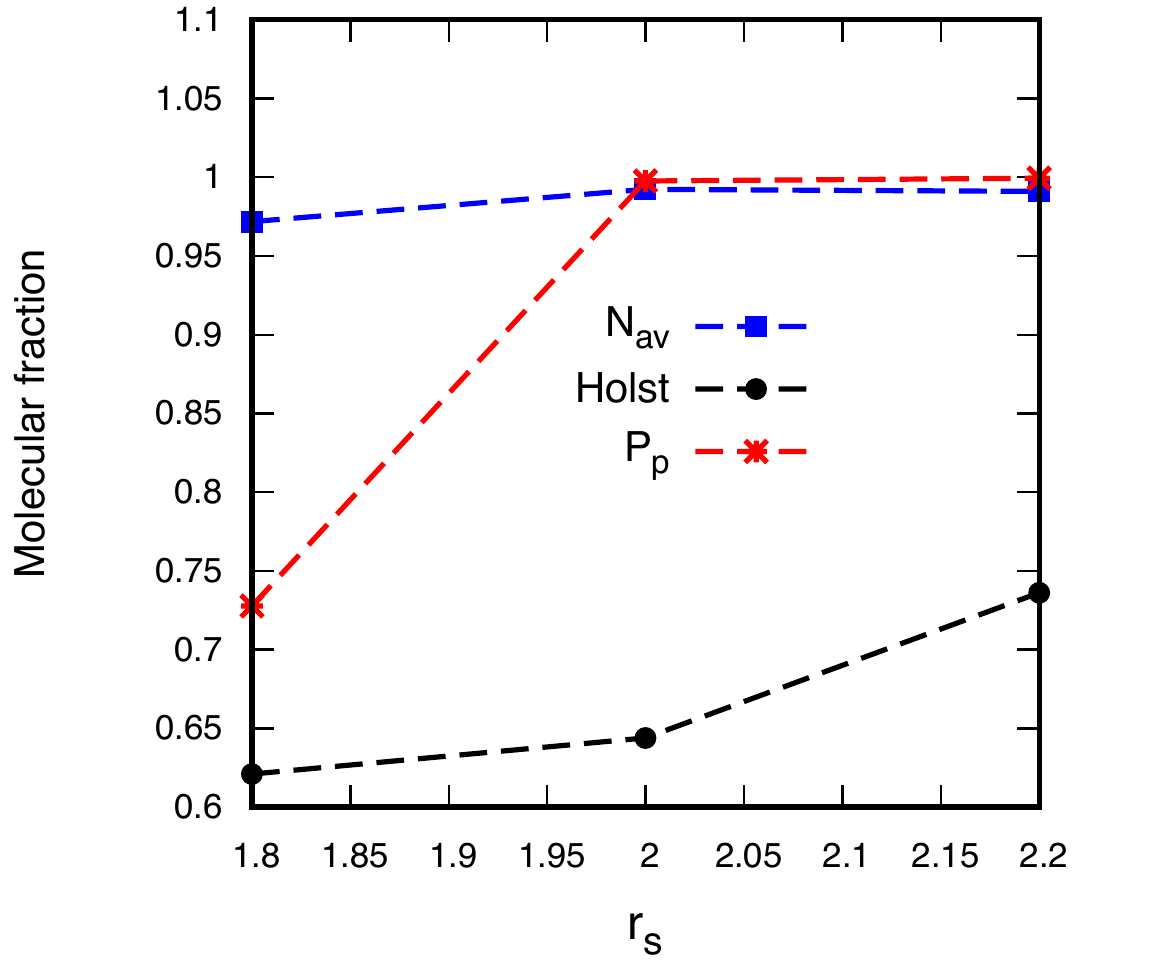}
\includegraphics[angle=0,width=\columnwidth]{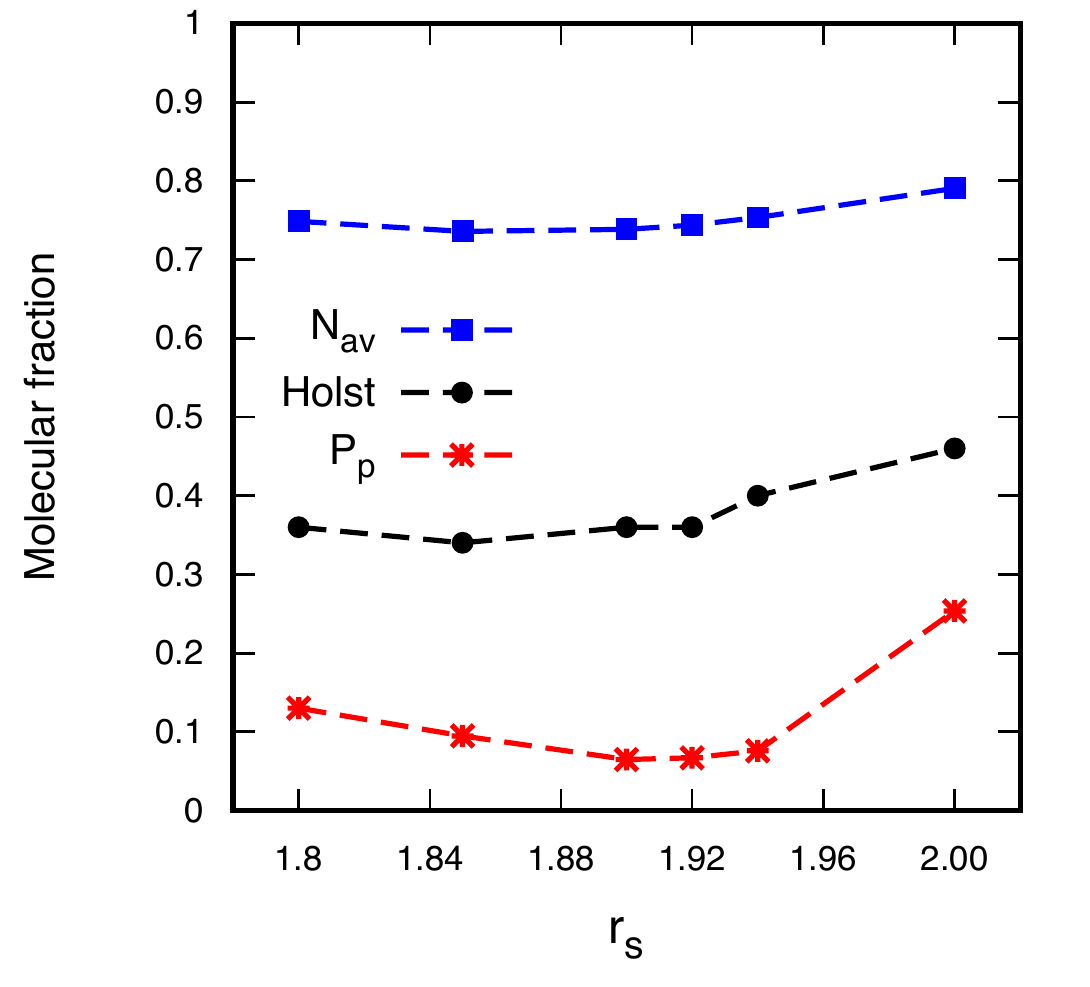}
\caption{Molecular fraction from three different estimators for the CEIMC trajectories along the T=4000K isotherm (upper panel) and the T=8000K isotherm (lower panel). $P_p$ is the probability for an atom to be paired to the same partner along the entire trajectory, $N_{av}$ the average number of pairs found within a cutoff distance corresponding to the first minimum of $g_{pp}(r)$ and {\it Holst} indicates the estimator based on the coordination number at the molecular peak of $g_{pp}(r)$.}
\label{fig:4TKKmfrac}
\end{figure}

In this appendix we provide some structural analysis along the three  isotherms investigated by CEIMC. We start by comparing CEIMC and DFT-BOMC proton-proton correlation functions at the densities reported in table \ref{tab:results}. Those comparisons are reported in figures \ref{fig:2TKKgr}, \ref{fig:4TKKgr} and \ref{fig:8TKKgr}.
In all cases we observe a rather good agreement between CEIMC and DFT-BOMC data confirming that the two models provide similar local environment, at least at the two-particle level, as already observed at higher pressure across the liquid-liquid transition line\cite{Gorelov2019}. Figure \ref{fig:2TKKgr} indicates the fully molecular character of the system along the $T=2000K$ isotherm in the investigated density range. At higher temperature we employed the cluster analysis detailed in Ref. \cite{Pierleoni2017} to compute the molecular fraction. Results at $T=4000K$ and $T=8000K$ are shown in figures \ref{fig:4TKKmfrac}. Different estimators for the molecular fraction has been proposed in the literature. In Ref. \cite{Pierleoni2017} we proposed an estimator based on the probability of an atom to be paired to the same partner along the entire simulation, called $P_p$, and we compared to two other estimators, the first one called $N_{av}$ is the average number of molecules found within a cutoff distance corresponding to the first minimum of $g_{pp}(r)$, and the second one is the coordination number at the first maximum of $g_{pp}(r)$ proposed in Ref. \cite{Holst2008}. Of the three estimators only $P_p$ implements the notion of persistence of bonding for well formed molecules and it was considered a better measure of the molecular fraction in particular in the dissociated regime. Here we observe that it is between the other two estimators at $T=4000K$ where $g_{pp}(r)$ has a strong molecular peak, while $P_p$ is rather lower than the other two estimators at $T=8000K$ where the molecular character of $g_{pp}(r)$ is only marginal. Moreover, as discussed in Ref. \cite{Pierleoni2017}, even this estimator in the dissociation regime can only be considered an upper bound to the molecular fraction. This analysis shows that the elbow in the Hugoniot, although related, is not determined by molecular dissociation because at $T=8000K$ at the Hugoniot conditions the molecular dissociation is almost exhausted.

\section{Sampling correction}
In the section we discuss the possible bias coming from nuclear configurations that have been sampled through a simplified procedure. This is indeed the case in our protocol where nuclear sampling is performed with a not-fully optimized VMC procedure and final results are obtained performing fully optimized VMC and RMC calculations for a set of fixed nuclear configurations. 

Let us consider the general problem of comparing the average of an observable $O(\bR)$ for two different models
with statistical weight $w_0$ and $w_1$, 
e.g. corresponding to
two different Born-Oppenheimer energy surfaces, $w_0 \sim \exp[- \beta V_0(\bR)]$ 
and $w_1 \sim \exp[-\beta V_1(\bR)]$, where $\beta$ is the inverse temperature. At thermal equilibrium we have 
\begin{equation}
    \langle \hat{O}\rangle_1=
    \frac{\int d\bR O(\bR) w_1}{\int d\bR w_1}=
    \frac{\int d\bR O(\bR) w_{10} w_0}{\int d\bR w_{10}w_0}=
    \frac{\langle O(\bR) w_{10}\rangle_0}{\langle w_{10}\rangle_0}
\end{equation}
where
\begin{equation}
    w_{10}=\frac{w_1}{w_0} = e^{-\beta[V_1(\bR)- V_0(\bR)]}
    \label{eq:weights}
\end{equation}
Stochastically sampling the configurations in continuum space within Monte Carlo methods
the averages are estimated by
\begin{equation}
    \langle \hat{O}\rangle_0\simeq \frac{\sum_iO(\bR_i)}{\mathcal{N}}
\end{equation}
where the set of configurations $\{\bR_i, i=1,\mathcal{N}\}$ is extracted with probability proportional to $w_0$. An unbiased estimate of averages for the other model hamiltonian
corresponding to weight $w_1$, is then obtained by reweighting
\begin{equation}
    \langle \hat{O}\rangle_1\simeq \frac{\sum_iO(\bR_i) w_{10}(\bR_i)}{\sum_i w_{10}(\bR_i)}
\end{equation}

\begin{figure}
\includegraphics[angle=0,width=\columnwidth]{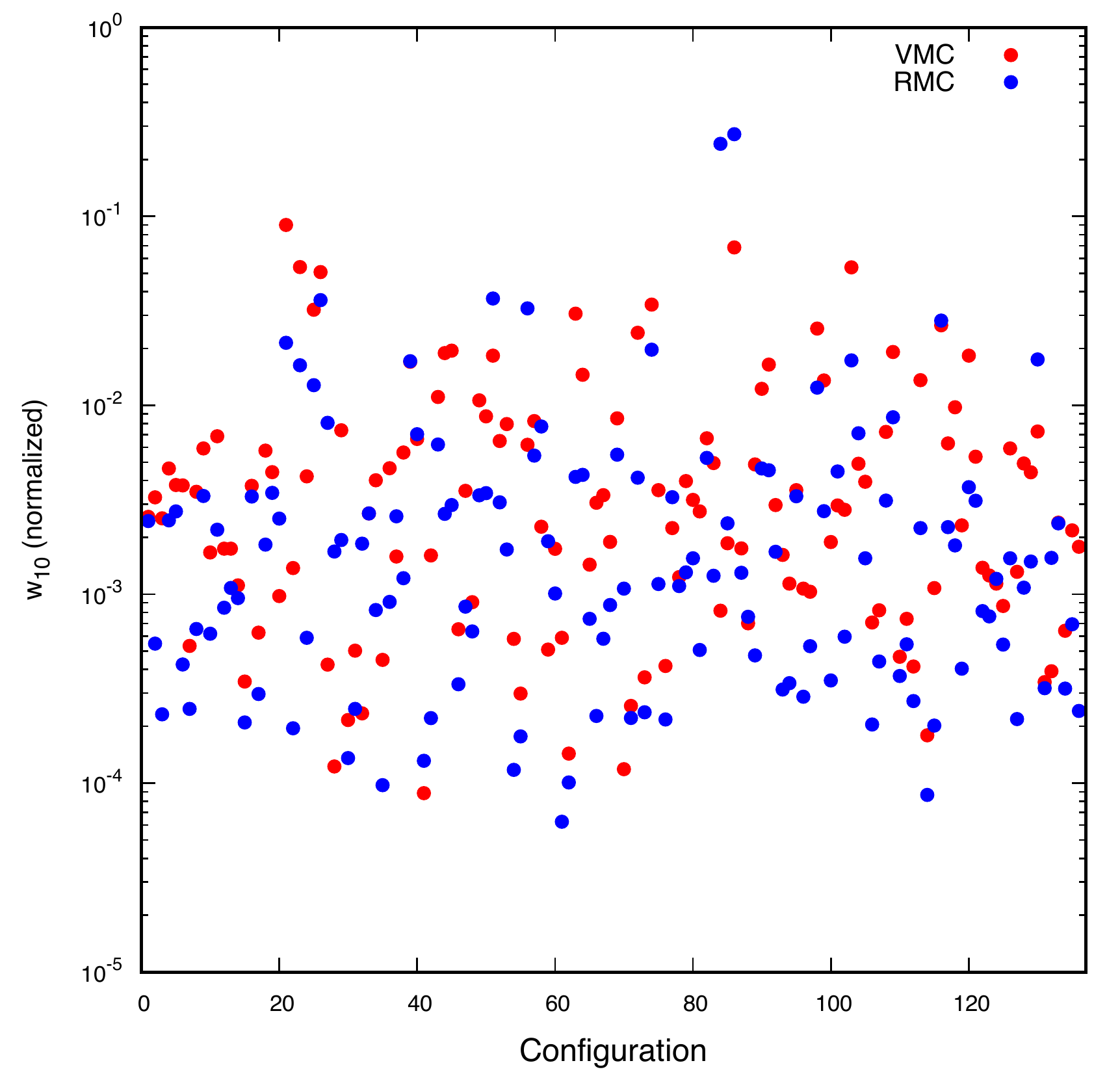}
\caption{Normalized, reweighted Boltzmann weight of 136 nuclear configurations for deuterium at $T=8000$ K and $r_s = 1.9$. We can see how the weights of the single configurations generated with CEIMC span several orders of magnitude.} 
\label{fig:norm_w}
\end{figure}

Although formally exact this procedure is not useful if the weights are wildly varying (note the extensive character of the exponent in eq. (\ref{eq:weights})): for a finite set of configurations only few will dominate the sums. 
This is the case when reweighting nuclear configurations generated during the CEIMC sampling with their proper weight either from fully optimized VMC (O-VMC) or from RMC. An example is given in figure \ref{fig:norm_w} where for 136 different nuclear configurations we report both the weights from O-VMC and RMC. 
It is however possible to introduce a cutoff to exclude outliers, i.e. configurations with a disproportionately large statistical weight, that would otherwise dominate the whole reweighted sampling.
By doing so we have recomputed energy, pressure and the hugoniot for two systems at $T=8000$ K, at $r_s = 1.85$ and 1.90. We show our results in table \ref{tab:reweighting}. We can see that even if there are a few small differences in the energies and pressures the results are largely compatible. Moreover we see that the estimates for the hugoniot are well within error bars. This means that CEIMC is able to provide a sample of configurations that can be used to accurately describe the system even after wave function optimization and projection in imaginary time, corroborating the overall robustness of our method.

\begin{table}[tpb]
\caption{Total energy, pressure and hugoniot averaged over 136 configuration for $T=8000$ K and $r_{S} = 1.85$ and $1.9$, obtained from arithmetic means and using reweighting.}
\label{tab:reweighting}
\begin{center}
\begin{tabular}{c|cc|cc}
\hline
$r_s = 1.85$ & VMC & Reweigth & RMC & Reweigth \\
\hline
$E$ (Ha/at.)  & -0.5085(2)  & -0.5081(3)  & -0.5120(2)  & -0.5114(3)  \\ 
$P$ (a.u.)& 0.001533(16)& 0.001550(17)& 0.001475(16)& 0.001518(17)\\ $H$ (Ha/at.)  & -0.0136(9)  & -0.0140(10) & -0.0083(9)  & -0.0100(10) \\ 
\hline
$r_s = 1.90$ & VMC & Reweigth & RMC & Reweigth \\
\hline
$E$ (Ha/at.)  & -0.5098(2)  & -0.5095(4)  & -0.5136(2)  & -0.5128(3) \\ 
$P$ (a.u.)& 0.001295(14)& 0.001318(19)& 0.001236(13)& 0.001266(18)\\ 
$H$ (Ha/at.)  & -0.0006(8)  & -0.0013(11) & 0.0044(7)   & 0.0036(11) \\ 
\hline
\end{tabular}
\end{center}
\end{table}

\section{Convergence of the RMC calculations}
\begin{figure*}
\includegraphics[width=\columnwidth]{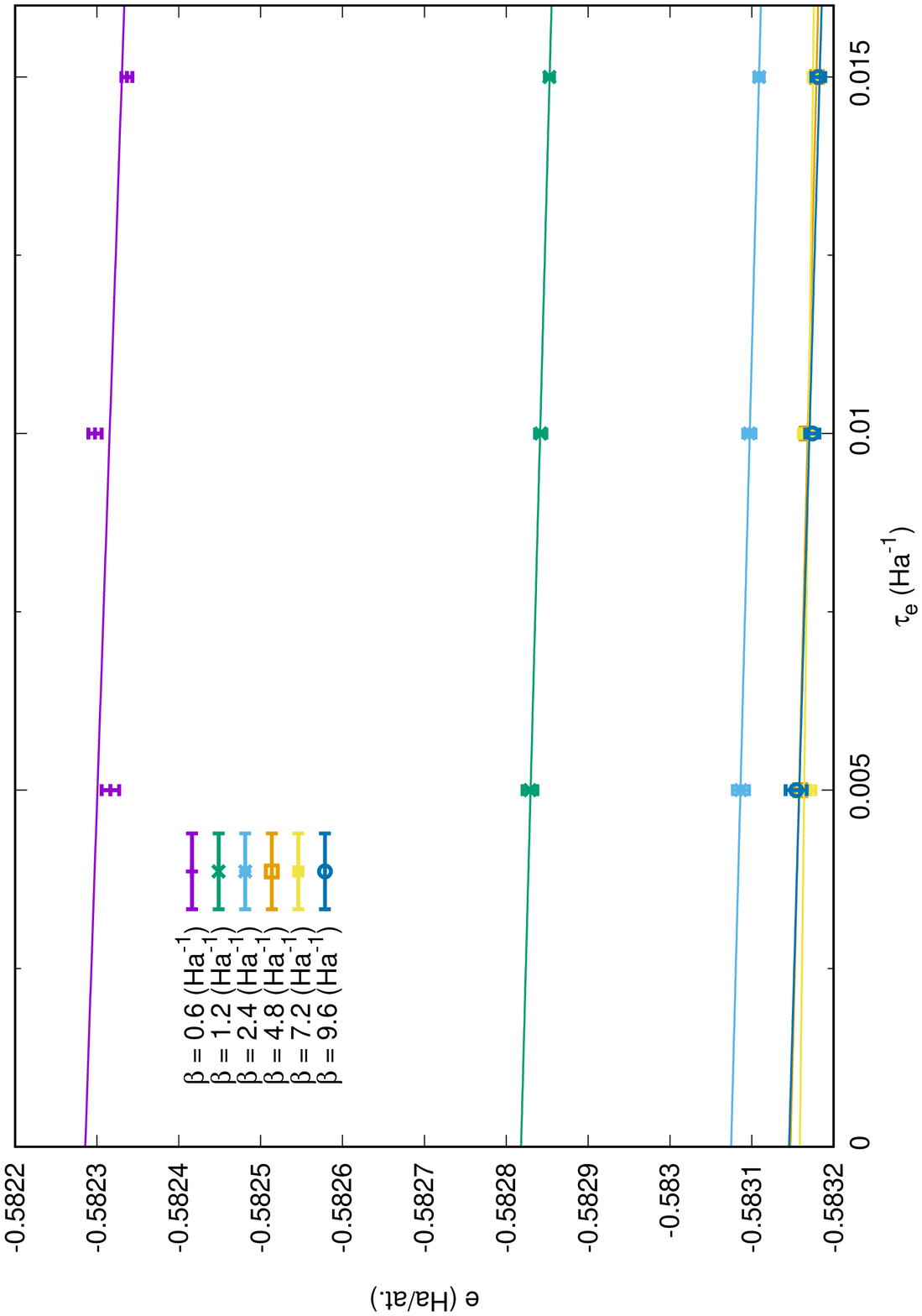}
\includegraphics[width=\columnwidth]{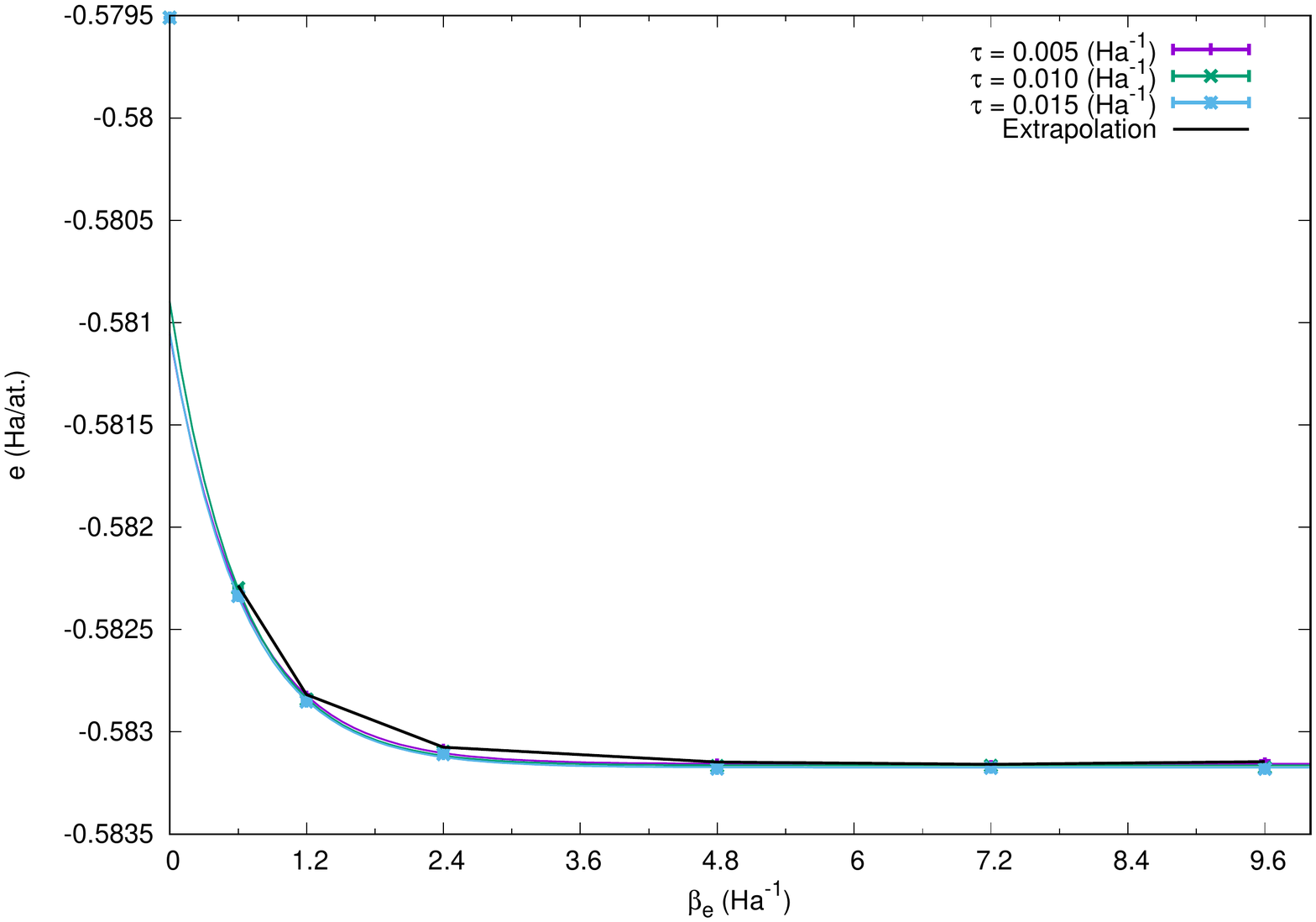}
\includegraphics[width=\columnwidth]{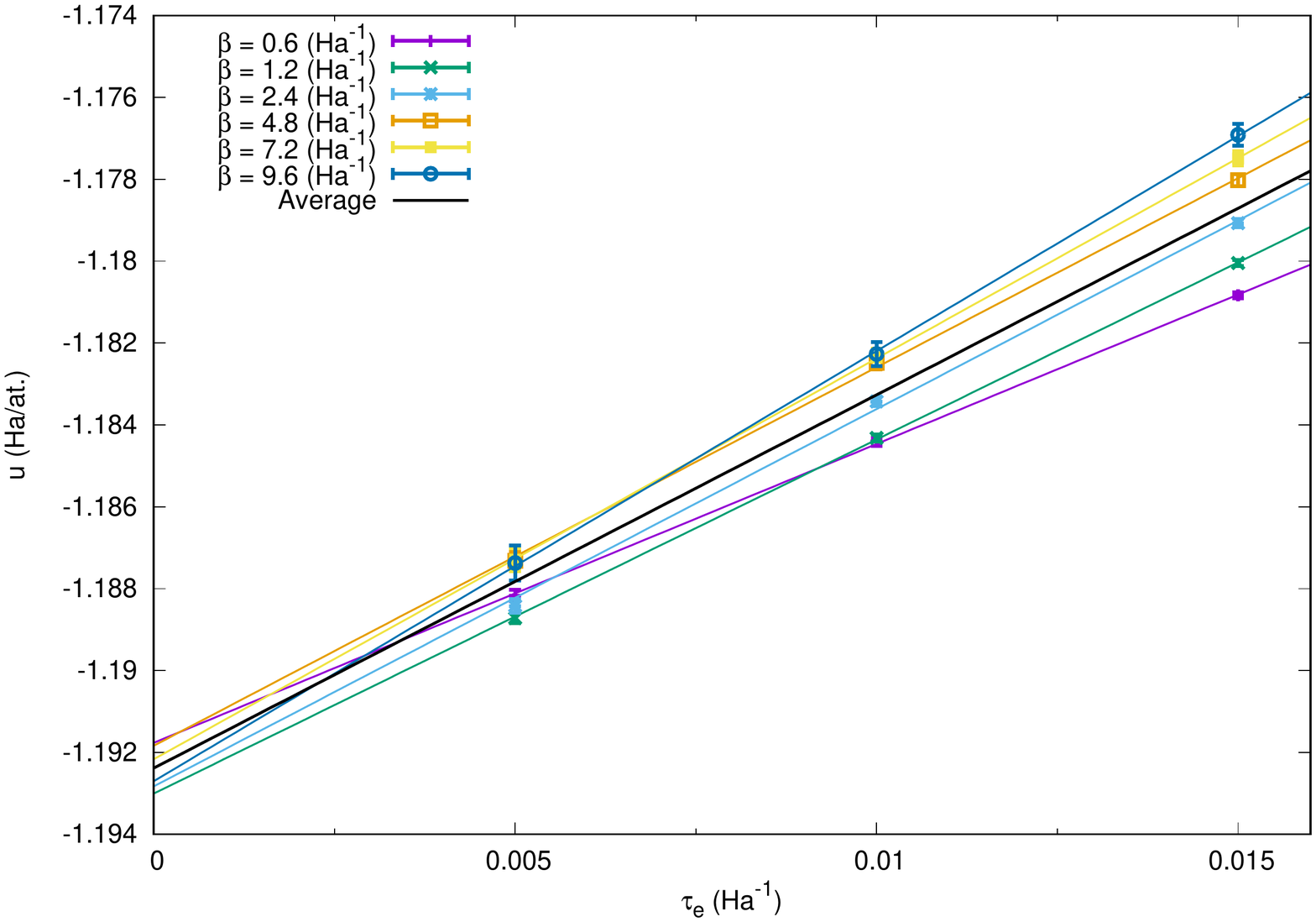}
\includegraphics[width=\columnwidth]{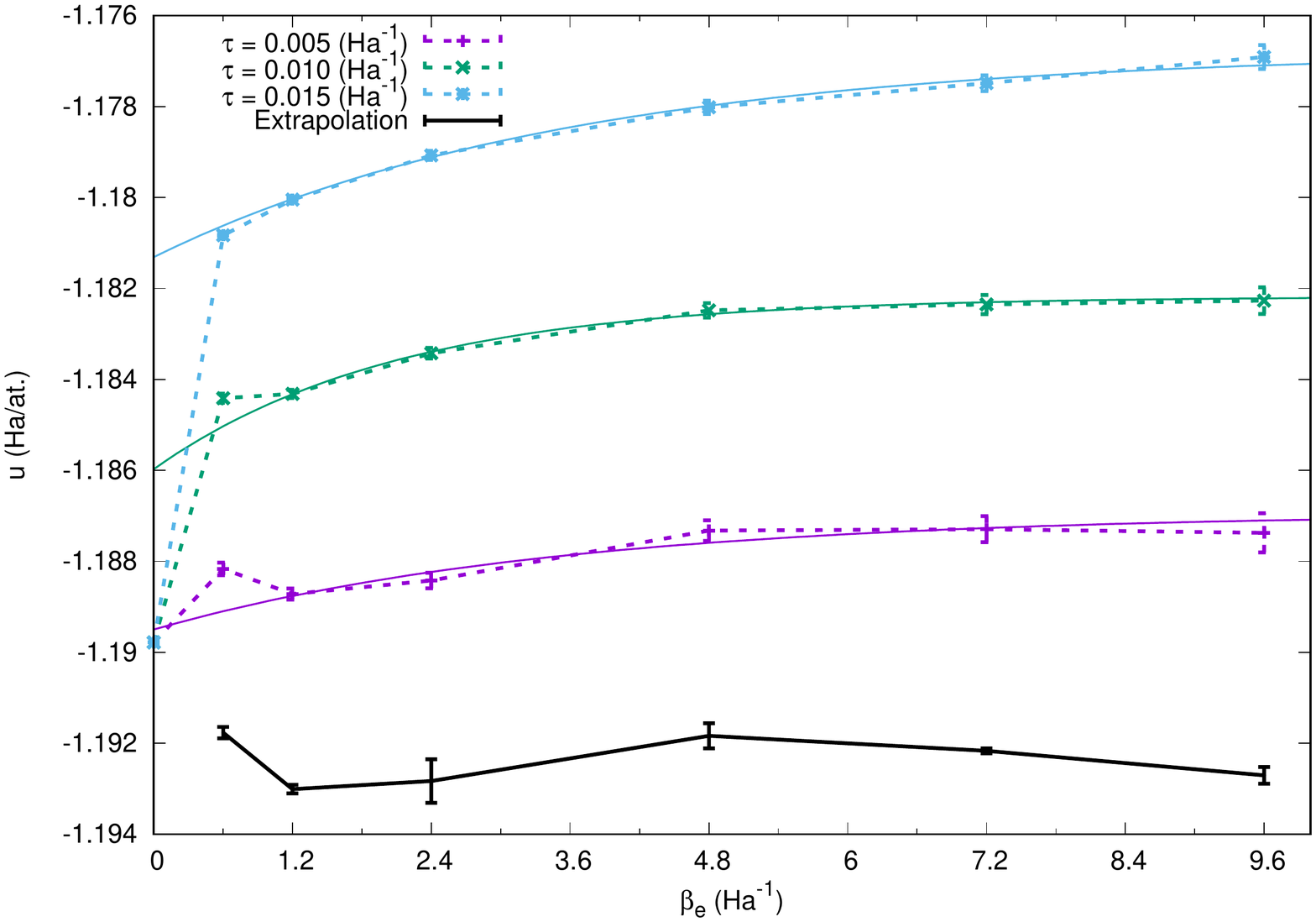}

\caption{RMC convergence with respect to $\tau_e$ and $\beta_e$ for total and potential energy.
} 
\label{fig:rqmc}
\end{figure*}

In this last section we report convergence tests for our RMC calculations.
In RMC results for finite imaginary time step ($\tau_e$) and finite projection in imaginary time ($\beta_e$) need to be extrapolated to the limits $\tau_e\to 0$, $\beta_e\to \infty$. In figure \ref{fig:rqmc} we report, for a single nuclear configurations, total and potential energy dependence on $\tau_e$ for various $\beta_e$ and on $\beta_e$ for various $\tau_e$. As for the total energy we see a very small dependence on $\tau_e$ and an exponential decaying behaviour of $e(\beta_e)$ as expected. Potential energy is more cumbersome. First we observe a strong $\tau_e$ linear dependence at fixed $\beta_e$ with the linear slope slightly depending on $\beta_e$. When plotting $u(\beta_e,\tau_e)$ vs $\beta_e$ for various $\tau_e$ we observe a growing exponential behaviour at large $\beta_e$ and a non-monotonous behavior at short $\beta_e$. This procedure needs to be performed for all twists and all different nuclear configurations before averaging. 
It is one the most delicate point of our procedure,
since the convergence of the potential energy propagates to the pressure estimator.
We note that the extrapolations introduce an important source of systematic bias of the  RMC results, notably for the pressure, which is absent
in the corresponding VMC
calculations.

\end{document}